\documentclass[10pt]{article}
\usepackage[OE]{express}

\newcommand{\ts}{\Delta}
\newcommand{\neff}{n_{\rm eff, MS}}
\newcommand{\neffP}{n_{\rm eff, p}}

\newcommand{\pmm}{{\rm PMM}}
\newcommand{\mic}{~\mu\rm m}

\begin{document}
\title{Multi-stage generation of extreme ultraviolet dispersive waves by tapering gas-filled hollow-core anti-resonant fibers}

\author{Md. Selim Habib\authormark{1,2}, Christos Markos\authormark{1}, J. Enrique Antonio-Lopez\authormark{2}, Rodrigo Amezcua Correa\authormark{2},  Ole Bang\authormark{1} and Morten Bache\authormark{1}}
\address{\authormark{1}DTU Fotonik, Technical University of Denmark, Kgs. Lyngby, DK-2800, Denmark\\
\authormark{2}CREOL, The College of Optics and Photonics, University of Central Florida, Orlando, FL-32816, USA}
\email{mdselim.habib@creol.ucf.edu}



\begin{abstract}
In this work, we numerically investigate an experimentally feasible design of a tapered Ne-filled hollow-core anti-resonant fiber and we report the generation of multiple dispersive waves (DWs) in the range 90-120 nm, well into the extreme ultraviolet (UV) region. The simulations assume a 800 nm pump pulse with 30 fs 10 $\mu$J pulse energy, launched into a 9 bar Ne-filled fiber with $34\mic$ initial core diameter that is then tapered to a $10\mic$  core diameter. The simulations were performed using a new model that provides a realistic description of both loss and dispersion of the resonant and anti-resonant spectral bands of the fiber, and also importantly includes the material loss of silica in the UV. We show that by first generating solitons that emit DWs in the far-UV region in the pre-taper section, optimization of the following taper structure can allow re-collision with the solitons and further up-conversion of the far-UV DWs to the extreme-UV with energies up to 190 nJ in the 90-120 nm range. This process provides a new way to generate light in the extreme-UV spectral range using relatively low gas pressure.
\end{abstract}

 % REPLACE WITH CORRECT OCIS CODES FOR YOUR ARTICLE, MINIMUM OF TWO; Avoid using the OCIS codes for “General” or “General science” whenever possible.
%For a complete list of OCIS codes, visit: https://www.osapublishing.org/oe/submit/ocis/
\bibliography{literature}
\bibliographystyle{osajnl}

\section{Introduction}

The VUV spectral range (10-200 nm) is of great technical and scientific interest because it is associated with an extensive range of applications, such as transient absorption spectroscopy \cite{Kobayashi2012}, ultrafast photochromic dynamics \cite{Tamai2000}, photoemission spectroscopy \cite{Reinert2005}, photoionization mass spectroscopy \cite{Hanna2009}, and control of chemical reactions \cite{Asplund2002}. %. However, the availability of femtosecond laser sources in that spectral region is extremely limited and they require a lot of space with bulk optics. Therefore, researchers around the globe have focused their research activities towards alternative solutions to develop tunable light sources suitable for VUV in a cost-effective manner.
However, the availability of femtosecond laser sources in that spectral region is scarce: standard frequency conversion of near-IR lasers in crystals is limited to the near-UV (300-400 nm) and mid-UV (200-300 nm), while in the far-UV (120-200 nm) and extreme-UV (XUV, 10-120 nm) the UV losses and lack of phase-matching conditions make harmonic conversion in crystals difficult or even impossible. In turn, harmonic generation in gas jets require high-energy pulses from low-repetition rate lasers and has low yield. 

Hollow-core anti-resonant (HC-AR) fibers \cite{Yu.HCARF.review.2016,Wei2017.HCARF.review,markos-RevModPhys.89.045003} are currently revolutionizing the fiber optics community. By structurally optimizing the cladding tubes surrounding the fiber core (shape, size, position, number), extremely low propagation loss can be achieved, the bend loss can be reduced, the fiber can be effectively made single-mode, and can have 40+ dB extinction ratio between the gas-filled core and the glass cladding \cite{Pryamikov2011.HCARF,Kolyadin2013.HCARF.nontouch,Yu:2012,Habib:2015.nested,Belardi2014.HCARF.nested,Belardi2014.HCARF,Poletti2014.HCARF.nested,Habib2016:ellipse,habib:2016.semicircle,Debord2017.HCARF.loss,Yu2016a.HCARF.experiment,Carter2017.HCARF.bendloss,Uebel2016.HCARF.HOM}. Especially the latter allows using the well-established telecommunication silica fiber production platform to achieve low-loss, broadband guidance in the important mid-IR range \cite{Pryamikov2011.HCARF,Kolyadin2013.HCARF.nontouch,Yu:2012,Habib:2015.nested}, and high-power delivery of pulsed lasers \cite{Michieletto2016.HCARF,urich.2013.HCARF.highpower,Jaworski2013.HCARF.power}. Compared to bulk solid-state matter, gas-filled HC-AR fibers also has remarkable properties for ultrafast nonlinear optics, such as extremely high damage threshold, tunable dispersion and nonlinearity through the gas pressure, and very long confinement lengths \cite{Travers:2011.JOSAB,Russell2014.HCPCF.review}. Finally, at high intensities ionization of the gas results in plasma formation \cite{holzer.2011.PhysRevLett.107.203901}, which because it is confined in a fiber is promising to spark a new era of filament-based nonlinear optics \cite{Ermolov-2015-PhysRevA.92.033821,kottig.2017.PhysRevLett.118.263902,saleh.plasma.2011.PhysRevA.84.063838,saleh.2011.plasma.PhysRevLett.109.113902,Habib2017,novoa.2015.PhysRevLett.115.033901}.

%The light-gas interaction and soliton dynamics in noble gas-filled HC-AR fibers have been extensively studied both in the near-IR \cite{Travers:2011.JOSAB,Russell2014.HCPCF.review,saleh.plasma.2011.PhysRevA.84.063838,saleh-PhysRevLett.107.203902,markos-RevModPhys.89.045003,Hasan2016} and mid-IR spectral regimes \cite{Habib2017,Hasan2016,Kottig2017-natcom}. 
Recent work has focused on using HC-AR fibers to solve the challenge of providing access to energetic ultrashort UV laser sources \cite{Im:2010,joly:2011,Chang2011a,Mak2013,Belli:2015-VUV-DW,Ermolov-2015-PhysRevA.92.033821,kottig.2017.PhysRevLett.118.263902,Kottig2017-natcom}.
Bright emission of mid-UV pulses using Ar-filled HC fiber was first predicted \cite{Im:2010} and later experimentally confirmed in \cite{joly:2011}. This relied on efficient generation of UV dispersive waves (DWs) by the well-known process of soliton self-compression to the sub single-cycle regime, followed by resonant DW radiation. Later work \cite{Belli:2015-VUV-DW,Ermolov-2015-PhysRevA.92.033821,kottig.2017.PhysRevLett.118.263902,Kottig2017-natcom} used the same approach to study further how to go deeper into the UV and achieving higher power 
UV light generation at 280 nm was also reported in a solid-core photonic crystal fiber (PCF) \cite{Stark2012}, however, the generation of VUV light was limited due to the strong material absorption in the VUV range.  

To date the work on VUV light generation in gas-filled HC-AR fibers has been carried out in uniform HC-AR fibers. In this work we numerically demonstrate (for the first time to the best of our knowledge) how the use of a tapered Ne-filled HC-AR fiber can lead in the first stage to the generation of multiple DWs in the far-UV region, which are then in the tapering section converted to XUV DWs through re-collision with the parent solitons. 
The process reaches XUV wavelengths down to 92 nm by tapering a $34\mic$ core diameter to $10\mic$. Because the tapering uses a two-step conversion a much lower gas pressure can be used compared to previous reports \cite{Ermolov-2015-PhysRevA.92.033821,Mak2013,Koettig2017}. We show how every section of the taper profile (transition length, waist, up-taper) has a crucial role in the soliton-plasma dynamics, which eventually directly affects the location and intensity of the generated DWs. In other words the taper profile should be carefully designed. In order to provide accurate numerical simulations of the continuous transitions of the fiber loss and dispersion resonances throughout the taper section, we had to develop a new analytical model. It is able to extend the simple capillary model to include loss and dispersion resonances without any significant assumptions. 

\section{Dispersion and loss of HC-AR fiber taper}

We seek to accurately model the XUV to near-IR nonlinear behavior of gas-filled HC-AR fibers. This is already a substantially complex task in a straight fiber due to the presence of a number of anti-resonant and resonant regions, affecting both losses and dispersion across the wavelength range. 
An additional challenge is that in the tapering section the fiber is scaled down in critical parameters (core size, thickness of cladding AR elements). Currently the accepted approach to model these fibers is to use the the so-called Marcatili-Schmeltzer (MS) capillary model in the nonlinear Schr\"odinger-like equation (NLSE) and neglect the resonances and how they affect the dispersion and loss. Only recently did some of us use the data from a full finite-element model (FEM) of the dispersion and loss and use it directly in an NLSE \cite{Habib2017,habib:2017.HCARF-UV-arxiv}. Recently also other groups investigated this by employing a Lorentzian extension of the MS model  \cite{Tani2018,Sollapur2017}. It is clear that accurate modeling of the resonances and the losses of these fibers is crucial, especially for predicting the UV behavior where the glass in the cladding becomes very lossy. 

\begin{figure}[t]
  \centering
  \includegraphics[width=\linewidth]{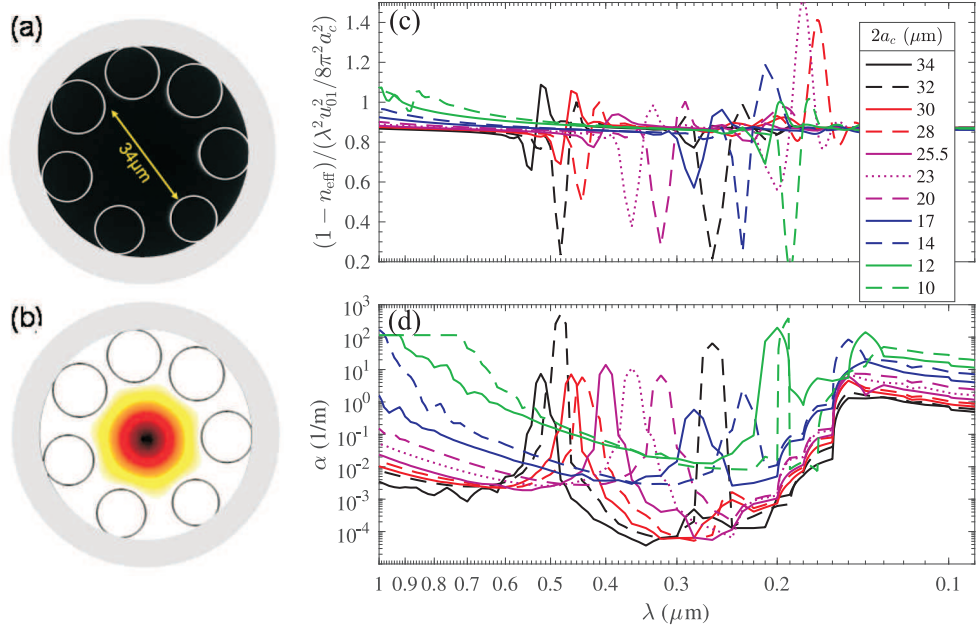}
\caption{(a) SEM image of a fabricated HC-AR silica fiber with 7 AR cladding tubes, 250 nm tube-wall thickness and $34\mic$ core diameter. (b) The FEM design used in this paper based on the fiber (a), overlapped with the fundamental mode at 800 nm. (c) and (d) FEM data of calculated effective index and loss vs. wavelength. The original fiber ($2a_c=34\mic$ and $\Delta=250$ nm) was used as a starting point, and the shown cores sizes are then considered linearly tapered so $\Delta$ is scaled accordingly down. }
%\vspace{-12pt}
\label{fig:fiber_loss}
\end{figure}

A scanning electron microscope (SEM) image of the HC-AR fiber used in our calculations is shown in Fig. \ref{fig:fiber_loss}(a). The HC-AR fiber has a core diameter of $34\mic$, an average capillary diameter of $16\mic$, and an average silica wall thickness of 250 nm. The  wall thickness was chosen to give a first AR transmission band centered at 800 nm. Figure \ref{fig:fiber_loss}(b) shows the near-field profile of the fundamental mode of the imported cross-section structure, calculated using FEM.  

It is important to mention that careful consideration of the silica refractive index is required for wavelengths less than 210 nm due to the absorption by impurities and the presence of OH groups and point defects \cite{Kitamura2007,tan2005:PhysRevB.72.205117}. In turn, the standard Sellmeier equation holds well in the range $\lambda>0.210\mic$ \cite{Kitamura2007,bache:2018}. Therefore, we used the standard Sellmeier equation \cite{ghosh:1994} for $\lambda>0.210\mic$ and the measured refractive index data of silica for $\lambda<0.210\mic$ to calculate the mode propagation constant $\beta$ and confinement loss $\alpha$ using the FEM-based COMSOL software. Importantly, we allowed the refractive index of silica to be complex in the FEM calculations, in order to account for the material losses in the UV. This approach is different than our previous work where the silica loss was not considered in the FEM calculation and the material-based loss was added afterwards based on the power-fraction of light in silica \cite{Habib:2015.nested}. 
To get an accurate calculation of the loss, we used a perfectly-matched layer outside the fiber domain and great care was taken to optimize both mesh size and the parameters of the perfectly-matched layer \cite{Habib:2015.nested,Poletti2014.HCARF.nested,habib:2016.semicircle}. 

\begin{figure}[t]
  \centering
  \includegraphics[width=\linewidth]{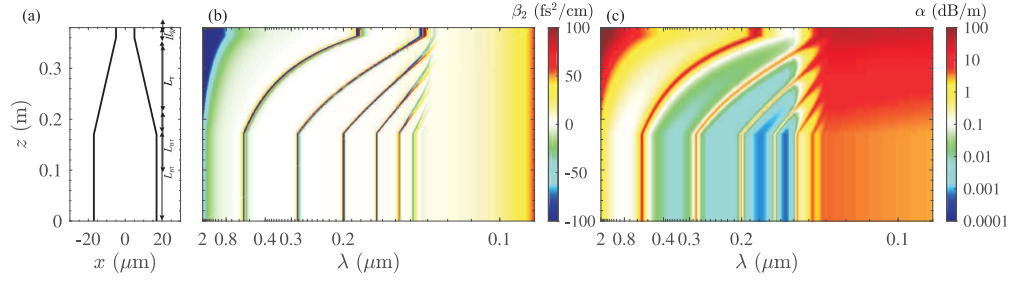}
\caption{(a) Basic fiber taper layout with $L_{\rm BT}= 17$ cm of untapered fiber with 34 $\mu$m core diameter followed by $L_{\rm T}= 13$ cm taper to a 10 $\mu$m core size and $L_{\rm W}=2$ cm waist after the taper. For a fiber filled with 9 bar Ne, (b) and (c) show the hybrid fundamental mode's GVD and loss vs. wavelength (log scale) using the poor-man's model \cite{bache:2018}, i.e. using  Eqs. (\ref{eq:neff_p})-(\ref{eq:alpha_total}). }
%\vspace{-12pt}
\label{fig:pmm}
\end{figure}

Figure \ref{fig:pmm}(a) shows the taper profile of the fiber used in our work with the optimum values of the length of the uniform waist section ($L_{\rm W}$), the down-taper section ($L_{\rm T}$), and the uniform input or before taper section ($L_{\rm BT}$). The core diameter of the $L_{\rm BT}$ and $L_{\rm W}$ sections was $34\mic$ and $10\mic$, respectively. The former is a quite typical value for 800 nm pumping, and can easily handle 10's of $\mu$J of pump energy. The latter waist value was chosen to give access to dispersive wave phase-matching deep into the XUV. 

In our model, we assume a linear decrease in silica wall thickness during the taper transition from 34 to $10\mic$ as just as the hole-to-hole pitch to a good approximation has been found to decrease linearly in tapered solid-core PCFs \cite{petersen:2017}. The silica wall thickness of $L_{\rm BT}$ is initially 250 nm, which then decreases to 74 nm in the taper waist section. This dramatically affects the resonance wavelengths, and thus both dispersion and loss of the fiber modes. We carefully modeled 11 different core sizes from 10 to $34 \mic$ and observed how the resonances changed. The aim was to seamlessly be able to track the resonances as they change through the taper due to a reduced core diameter and wall thickness of the cladding tubes. The result is shown in Fig. \ref{fig:fiber_loss}(c) and (d). 

However, implementing these data sets in the NLSE it turned out that much more than 11 FEM data sets are needed to do this since the resonance wavelengths significantly change from one core size to the next; this made it almost impossible to do a smooth interpolation during the tapering section. We therefore decided to investigate various strategies for a more analytical approach that could allow us to calculate the resonances for any core size and cladding tube wall thickness \cite{bache:2018}. We developed a so-called poor-man's model as an analytical extension of the celebrated MS capillary model. It can accurately describe the full  dispersion and loss profiles of AR-HC fibers, seamlessly covering all resonance and anti-resonance regions. Being analytical in nature it works for any core size, cladding wall thickness and glass-cladding material. All the model needs is a single quite general fitting parameter to match the overall loss level to a single FEM simulation, which serves as a standard for the chosen HC-AR fiber design (e.g. circular or elliptical cladding tubes, nested or no nested cladding tubes etc.). We refer to \cite{bache:2018} for more details, but briefly 
we start with the MS model and calculate the core modes of an evacuated capillary having a core radius  $a_c$ and a capillary thickness  $\Delta$.
The core modes are Bessel functions of the first kind $\propto J_m(\kappa r)$. Here we have introduced the core transverse wavenumber $
\kappa=k_0 \sqrt{1-\neff^2}
$ 
where $k_0=\omega_0/c=2\pi/\lambda_0$ is the vacuum wavenumber. Here $\neff$ is the effective index of the mode, as calculated from the eigenvalue problem, and in the specific case of an evacuated fiber under the perfect-conductor assumption, yielding $\kappa a_c=u_{mn}$, we have the simple relation \begin{align}
\neff^2=1-\frac{u_{mn}^2}{a_c^2k_0^2}
\end{align}
where $u_{mn}$ is the $n$'th zero of the $m$'th order Bessel function $J_m$.
In the dielectric the transverse wavenumber is 
$\sigma=k_0\sqrt{n_d^2-\neff^2}
$where $n_d$ is the dielectric refractive index. Since for a hollow fiber $\neff\simeq 1$, irrespective of whether the fiber is evacuated or gas-filled, we can to a good approximation write $\sigma\simeq k_0\sqrt{n_d^2-1}$. 
We may analytically calculate the mode loss in this thin capillary. This is done with a bouncing-ray (BR) model, and the loss coefficients for the TE and TM modes are then \cite{bache:2018}
\begin{align}
\label{eq:alphaTE}
\alpha_{\rm TE,BR}&=\frac{2u_{mn}}{a_c^2k_0 [
4\cos^2(\sigma \ts)+(\tfrac{\sigma}{\kappa}+\tfrac{\kappa}{\sigma})^2 \sin^2(\sigma \ts)]}
\\
\label{eq:alphaTM}
\alpha_{\rm TM,BR}&=\frac{2u_{mn}}{a_c^2k_0 [
4\cos^2(\sigma \ts)+(\tfrac{\sigma}{n_d^2\kappa}+\tfrac{n_d^2\kappa}{\sigma})^2 \sin^2(\sigma \ts)]}
\end{align}
For hybrid modes, including the fundamental mode with $m=0$ and found by taking the first zero $n=1$, the loss is taken as a geometric average. 
\begin{align}
\alpha_{\rm H,BR}&=(\alpha_{\rm TE,BR}+\alpha_{\rm TM,BR})/2
\label{eqn:alpha_hybrid}
\end{align}
We remark that this is as far as we know the first analytical direct calculation of the HC-AR fiber loss without resorting to perturbative methods. As we discuss in detail in \cite{bache:2018} alternative, perturbative methods \cite{Miyagi:1984,Archambault:1993,Zeisberger2017} do not give correct values at the resonance loss peaks.

% We also adopt the perfect-conductor approximation, which assumes that the guided core modes are zero at the core-dielectric interface. The radial nature of these modes are Bessel functions of the first kind $\propto J_m(\kappa r)$, yielding $J_m(\kappa a_c)=0$ under the perfect-conductor approximation. This implies that 
% \begin{align}
% \kappa a_c=u_{mn}
% \end{align} 

Next we take the basis in this loss calculation to perturbatively calculate the associated dispersion. Such a pertubative method for calculating the loss and dispersion was actually shown in the original MS model for an infinitely thick capillary \cite{Marcatili:1964}. It turns out \cite{bache:2018} that this can be generalized to the thin capillary to give the following perturbative extension of the effective index and power loss coefficient for a hybrid mode
\begin{align}
\label{eq:neff_p}
n_{\rm eff,p} &= \neff-\frac{u_{mn}^2}{a_c^3k_0^3}{\rm Im}(Z_{\rm H})
\\
\label{eq:alpha_p}
\alpha_{\rm H,p}&=\frac{2u_{mn}^2}{a_c^3k_0^2}{\rm Re}(Z_{\rm H})
\end{align}
where $Z_{\rm H}=(Z_{\rm TE}+Y_{\rm TM})/2$ is the hybrid mode impedance, calculated as an average of the TE impedance and the TM admittance. The critical next step is to get an accurate expression of the impedance. The analytical impedance expressions (also found in, e.g., \cite{Miyagi:1984,Archambault:1993}) yield a too high loss at the AR resonances, basically because the perturbative approach breaks down exactly around the resonances. In contrast, the BR model has no restrictions, and Eq. (\ref{eqn:alpha_hybrid}) must hold. Therefore to get the perturbative loss to match that of the BR model, $\alpha_{\rm H,BR}=\alpha_{\rm H,p}$, we suggested using the following modified impedances \cite{bache:2018}
\begin{align}
\label{eq:ZTE_mod}
\hat Z_{\rm TE}=Z_0\frac{\frac{1}{2}-i(\frac{\sigma}{\kappa}+\frac{\kappa}{\sigma})^{-1} \tan(\sigma \ts)}
{2-i(\frac{\sigma}{\kappa}+\frac{\kappa}{\sigma}) \tan(\sigma \ts)}
\\
\label{eq:YTM_mod}
\hat Y_{\rm TM}=Y_0\frac{\frac{1}{2}-i(\frac{\sigma}{n_d^2\kappa}+\frac{n_d^2\kappa}{\sigma})^{-1} \tan(\sigma \ts)}
{2-i(\frac{\sigma}{n_d^2\kappa}+\frac{n_d^2\kappa}{\sigma}) \tan(\sigma \ts)}
\end{align}
where $Z_0=k_0/\kappa$, $Y_0=n_0^2Z_0=Z_0$ (where $n_0=1$ is the core refractive index) and the hat denotes that the equations have been empirically  modified to match the loss of the BR model. 

Finally, in \cite{bache:2018} we also showed how the mode losses can be calculated using this model if the  dielectric is lossy $\bar n_d=n_d+i\tilde n_d$. Based on work by \cite{Archambault:1993}, we found that we must replace the transverse wavenumber ratio $\sigma/\kappa$ %in Eq. (\ref{eq:ZTE}) 
for the TE case with
\begin{align}
\label{eq:sigmakappa_lossy_TE}
\left(\frac{\sigma}{\kappa}\right)^*=
\frac{\sigma}{\kappa} \frac{1+\frac{\kappa}{\sigma}\tanh(n_d \tilde n_d Z_d^2\sigma\ts )}{1+\frac{\sigma}{\kappa}\tanh(n_d \tilde n_d Z_d^2\sigma\ts )}
\end{align}
where $Z_d=(n_d^2-1)^{-1/2}$. 
%In the same spirit the ratio $Z_0/Z_d$ in Eq. (\ref{eq:alphaTE}) should be substituted with the above expression. 
Similarly for the TM case %in Eq. (\ref{eq:YTM}) 
we must replace $\sigma/(n_d^2\kappa)$ with
\begin{align}
\label{eq:sigmakappa_lossy_TM}
\left(\frac{\sigma}{n_d^2\kappa}\right)^*=
\frac{\sigma}{n_d^2\kappa} \frac{1+\frac{n_d^2\kappa}{\sigma}\tanh(n_d \tilde n_d Z_d^2\sigma\ts )}{1+\frac{\sigma}{n_d^2\kappa}\tanh(n_d \tilde n_d Z_d^2\sigma\ts )}
\end{align}
%and the ratio $Y_0/Y_d$ in Eq. (\ref{eq:alphaTM}) should be substituted with the above expression. 
With these extensions we can evaluate analytically the expected mode loss due to the UV material loss in silica. In addition, the model directly gives the associated dispersion resonances through Eq. (\ref{eq:neff_p}).

To complete the poor-man's model for the losses we include an overall adjustment factor $f_{\rm FEM}$ that allows us to adjust the calculated spectral loss shape to match the levels found in the FEM data, so the total loss is 
\begin{align}
\label{eq:alpha_total}
\alpha_\pmm=f_{\rm FEM}\alpha_{\rm capillary}
\end{align}
where $\alpha_{\rm capillary}$ is the calculated capillary loss of the mode, i.e. either TE, TM or hybrid. In the standard case we calculate the fundamental hybrid mode, which relies on the eigenvalue determined by the first zero of the 0'th order Bessel function of the first kind, $u_{01}\simeq 2.405$. 
We found that $f_{\rm FEM}=10^{-2}$ gave good agreement with the FEM data of the fundamental hybrid mode at various core sizes of the fiber shown in Fig. \ref{fig:fiber_loss}(a). This choice was a compromise between matching the loss in the anti-resonant regions and the UV loss plateau \cite{bache:2018}. 

Having this semi-analytical model of the resonant and anti-resonant behavior of the HC-AR fiber, consisting of Eqs. (\ref{eq:neff_p})-(\ref{eq:alpha_total}), we can now get a complete picture of how the fiber properties change during the tapering section. 
Fig. \ref{fig:pmm}(b) and (c) show the dramatic changes in the numerous resonances along the fiber taper section: the first resonance moves from 550 nm to 180 nm and the second resonance from 280 nm to 135 nm. The model captures all resonances, and an important point is that in the UV the resonances eventually disappear. This is because we include the imaginary part of the refractive index into the model. We used the measured refractive index data of silica from \cite{Kitamura2007} to evaluate the complex refractive index for wavelengths from the XUV to the mid-IR. In the absence of material losses, when the core is large ($a_ck_0\gg 1$) we have $\sigma/\kappa\gg 1$, which gives the characteristic dramatic changes from resonant to anti-resonant losses as the wavelength is changed; basically in this limit the $\sin^2(\sigma\ts)$ is cyclically zero (resonance) or very large (anti-resonance) compared to the competing $\cos^2(\sigma\ts)$ term in the denominator of the losses Eqs. (\ref{eq:alphaTE})-(\ref{eq:alphaTM}). Instead when $\tilde n_d$ becomes significant, it turns out that $(\sigma/\kappa)^*\rightarrow 1$ (TE case) and  $(\sigma/n_d^2\kappa)^*\rightarrow 1$ (TM case). This implies that the modified impedances become constant and thus no resonances appear in the loss and dispersion. Specifically, in the loss equations in the denominators the $\cos^2(\sigma\ts)$ and $\sin^2(\sigma\ts)$ term now have the same prefactors, so they add up to unity, thus canceling the resonant/anti-resonant behavior.  %The loss extracted from the poor-man's model is very consistent with FEM data, see \cite{bache:2018}, because it includes both waveguide loss and material loss. 
Some interesting observations can be made: the XUV loss is rather constant during the taper, which is mainly because material loss affect the mode losses. It does increase to a critical level at the taper waist, $>10~\rm dB/m$. It is also worth noticing that the pump wavelength experiences very high losses at the end of the taper since it is now at the IR edge of the first anti-resonance band. Therefore the tapered fiber is not suitable for long propagation of neither the generated solitons nor the generated UV light.  

%Figure 1(d) shows the refractive index of silica used in our calculations, based on the Sellmeier equation in [41], which is valid from 0.21-3.7 μm (see red curve in Fig. 1(d)). However, it is important to mention that careful consideration of the silica refractive index is required for wavelengths less than 210 nm due to the absorption by impurities and the presence of OH groups and point defects [42,43]. Therefore, we used the measured refractive index data of silica from  [42] (see black curve in Fig. 1(d)) to calculate the group velocity dispersion (GVD) and confinement loss for wavelengths less than 210 nm. The GVD and loss of the evacuated HC-AR fiber for different core diameters is shown in Fig. 1(e-f). The GVD and loss was calculated using the finite-element based COMSOL software. To accurately calculate the loss, we used a perfectly-matched layer outside the fiber domain and great care was taken to optimize both mesh size and the parameters of the perfectly-matched layer [11,13,15]. 

%The resonance bands seen in the GVD profile in Fig. 1(e) are due to the resonant coupling between the core and cladding modes. These resonances also strongly affect loss, resulting in the localized peaks in the loss curve seen in Fig. 1 (f). The location of the resonance bands is mainly determined by the silica wall thickness of the fiber. For a core diameter of $34\mic$ and a silica wall thickness of 250 nm, the corresponding resonance band is located at 500 nm (red curve in Figs. 1 (e-f)).

\section{Nonlinear pulse propagation model} 
The nonlinear optical pulse propagation was studied using a generalized nonlinear Schr\"odinger equation which also accounts free-electron effects described as \cite{Travers:2011.JOSAB,saleh-PhysRevLett.107.203902,fedotoc-PhysRevA.76.053811}: 
\begin{align}
\left( {i{\partial _z} + \hat D (i{\partial _t}) + i\frac{\alpha_\pmm }{2} + \gamma(1+i\omega_0^{-1}\partial _t) {|A|}^2  - \frac{{\omega_p^2(z,t)}}{{2{\omega _0}c}} + i\frac{{ A_{\rm eff} {I_p}{\partial _t}{N_e}}}{{2 {|A|}^2 }}} \right)A = 0
\end{align}
where $A$ is the complex field envelope centered around the central angular frequency  $\omega_0$, $t$ is the time in the reference frame moving with the pump group velocity, $\hat D (i\partial_t)=\sum_{m\geq 2}\beta_m (i\partial_t)^m/m!$ is the full dispersion operator in time domain and $\beta_m$ is the m$^{\rm th}$ order dispersion coefficient. 
The linear propagation constant of the gas-filled fiber was calculated using the following expression:
\begin{align}
\beta (\lambda ,p,T) = \frac{{2\pi }}{\lambda }\sqrt{\neffP^2(\lambda ) + \delta (\lambda )\frac{\rho (p,T)}{\rho _0}}
\end{align}
where $\neffP (\lambda)$ is effective mode index in an evacuated fiber, as calculated with the poor-man's model, $\delta(\lambda)$ is the Sellmeier terms describing the gas deviation from vacuum 
\cite{Borzsonyi:2008}, $\rho_0$ is the density of the material at standard temperature and pressure, and $\rho$ is the density at pressure $p$ and temperature $T$. The linear propagation loss of the fiber $\alpha_\pmm$ does not depend on pressure, and was calculated using the poor-man's model. 

As usual $\gamma=\omega_0 n_2/(cA_{\rm eff}) $ is the fiber Kerr nonlinearity, where $c$ is the velocity of light in vacuum, $A_{\rm eff}$ is the effective mode area of the fiber at the pump wavelength, and $n_2$ is the Kerr nonlinear coefficient of the gas, which scales linearly with pressure \cite{Borzsonyi:2010}. Self-steepening is included through the term $(1+i\omega_0^{-1}\partial _t)$. 
Furthermore, $\omega_p(z,t)=e\sqrt{N_e(z,t)/\varepsilon_0m_e}$ is the plasma frequency, $m_e$ is the electron mass,  $\varepsilon_0$ is the free space permittivity, $I_p$ is the ionization energy of the gas, and $N_e(z,t)$ is the free electron density. 

We neglect the Raman contribution of silica due to the very low light-glass overlap ($\ll0.1\%$) \cite{Chang2011a}. The dynamics of ionization depends on the multiphoton or tunneling process which is determined by the Keldysh parameter \cite{saleh-PhysRevLett.107.203902,saleh.plasma.2011.PhysRevA.84.063838}. In this work, the peak intensity at the maximum compression point reaches 350 TW/cm$^2$ and tunneling ionization dominates over multiphoton ionization. The free electron density was calculated using quasi-static tunneling ionization based on the Ammosov, Delone, and Krainov model \cite{saleh.plasma.2011.PhysRevA.84.063838,Chang2011a}. The free electron density was calculated with \cite{Chang2011a} $N_e (t) = N_0 \left( 
{1 - \exp \left[ - \int_{ - \infty }^t {W(t' )} {dt}' \right]} 
\right)$ 
% \begin{align}
% N_e (t) = N_0 \left( 
% {1 - \exp \left( - \int_{ - \infty }^t {W(t' )} {dt}' \right)} 
% \right),
% \end{align}
where $W(t)$ is the ionization rate, which was also calculated according to \cite{Chang2011a}.

\section{Numerical results}

\begin{figure}[t]
  \centering
  \includegraphics[width=\linewidth]{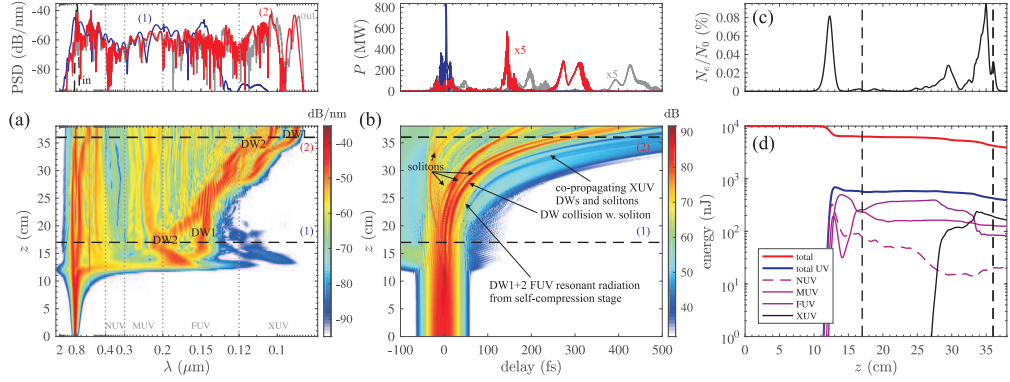}
\caption{Simulation of fiber taper shown in Fig. \ref{fig:pmm} using a 800 nm 30 fs $10~\mu$J input pulse. (a) Spectral evolution  has the NUV, MUV, FUV and XUV sections indicated with dotted lines, and the start (1) and finish (2) of the tapering section are indicated with dashed lines, and their snapshots are plotted with blue and red colors in the top plot. (b) Time evolution, shown with a dB scale to emphasize the DW dynamic; the snapshots at the end of the tapering section and fiber exit are amplified by a factor of 5 for better visualization. (c) Ionization fraction and (d) energy content in the various relevant sections of the spectrum vs. propagation distance.}
%\vspace{-12pt}
\label{fig:PSD}
\end{figure}

In the simulations we model an 800 nm 30 fs FWHM pulse with 10 $\mu$J energy and 1 kHz repetition rate. The fiber is filled with 9 bar Ne; the pressure is high enough to get soliton pulse compression after around 10 cm with a reasonable pulse energy, but it is much lower than, e.g. \cite{Ermolov-2015-PhysRevA.92.033821}. Ne was chosen as lighter gases like He and Ne are more transparent in the FUV and XUV than heavier gases like Ar. 

\begin{figure}[t]
  \centering
  \includegraphics[width=\linewidth]{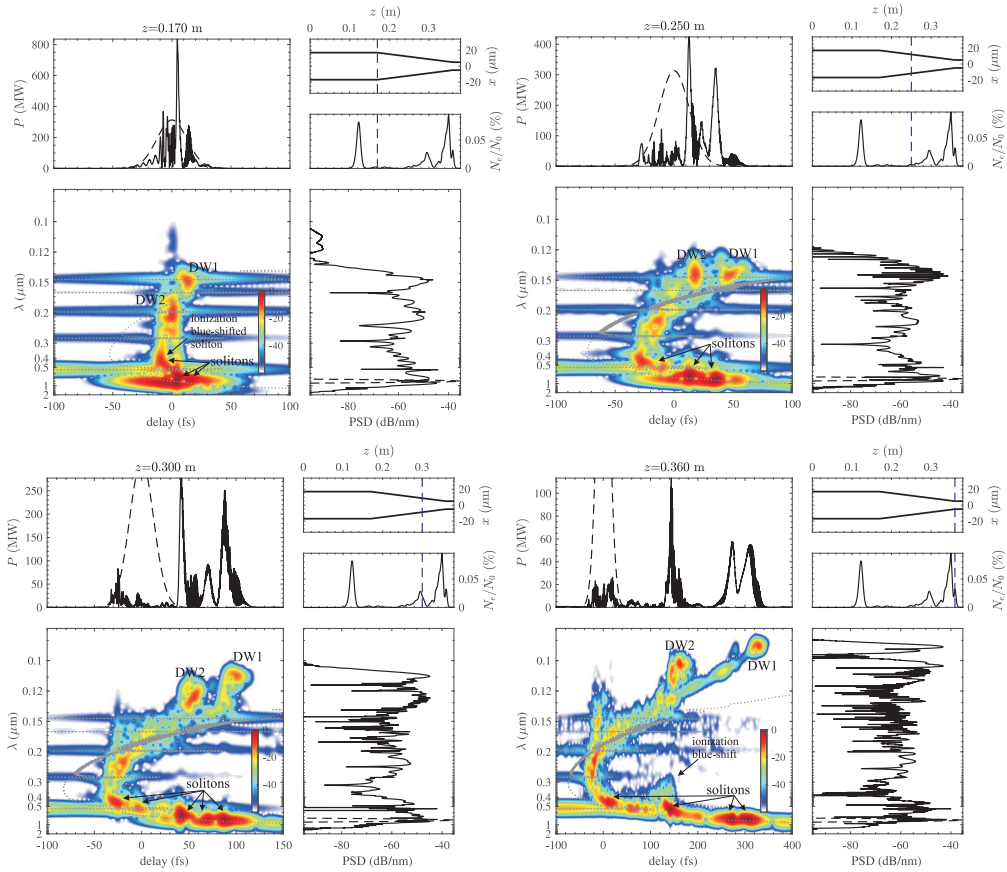}
\caption{Spectrograms at selected stages of the simulation in Fig. \ref{fig:PSD}. A 5 fs FWHM gating pulse was used. The dotted gray line shows the accumulated dispersion delay. On top of the spectrogram the power time trace is shown, while to the right the PSD spectrum is shown. In the top-right corner we show in each spectrogram the fiber taper shape and the ionization fraction vs. distance, and the current position is marked with a blue dashed line.}
%\vspace{-12pt}
\label{fig:XFROG}
\end{figure}

Fig. \ref{fig:PSD} shows a typical simulation where the untapered section is long enough for a soliton self-compression stage to form, which happens after 12 cm. This initial stage therefore gives us an indication of what kind of dynamics we would observe in a standard untapered fiber. At the self-compression stage, a sub-1 fs self-compressed soliton forms stretching from the near-IR to the FUV. In this process a strong FUV dispersive wave with $0.6~\mu$J total energy, centered at 170 nm. This wavelength is consistent with phase-matching calculations of degenerate four-wave mixing between the self-compressed soliton and the DW \cite{novoa.2015.PhysRevLett.115.033901}: $\beta_{\rm sol}(\omega_{\rm RR})=\beta(\omega_{\rm RR})$, where  $\omega_{\rm RR}$ is the resonant radiation frequency of the DW. Here $\beta_{\rm sol}(\omega)=\beta(\omega_0)+(\omega-\omega_0)\beta_1+\gamma P_{\rm sol}\omega/\omega_{\rm sol}-\frac{N_e}{2N_c}\frac{\omega_{\rm sol}^2}{c\omega}$ is the soliton dispersion relation including the Kerr nonlinear phase shift $\gamma P_{\rm sol}\omega/\omega_{\rm sol}$, where $P_{\rm sol} $ is the soliton peak power and $\omega/\omega_{\rm sol}$ accounts for self-steepening, and $-\frac{N_e}{2N_c}\frac{\omega_{\rm sol}^2}{c\omega}$ is the self-defocusing nonlinear plasma phase shift, where $N_c=\varepsilon_0m_e\omega_0^2/e^2$ is the critical plasma density where it becomes opaque \cite{Couairon:2007}. As the self-compressed soliton forms, it ionizes the gas partially, and this leads to a significant loss in the total energy. This ionization event is accompanied by soliton fission and it blue-shifts one of the solitons to below 500 nm, as we also will see later in the spectrograms in Fig. \ref{fig:XFROG}, and this actually red-shifts the first DW (175 nm) compared to what we would expect from a 800 nm soliton (120 nm). The DW breaks quickly into two, labeled DW1 and DW2, and following this soliton fission occurs resulting in 4 solitons, and this is the situation as the start of the tapering section arrives. 

If the fiber is not tapered the UV DWs would have a slower group velocity than the soliton, and because the gas is not Raman active no red-shifting and consequently no slowing-down of the solitons occurs. Therefore the solitons would never encounter the DWs again. However, in the tapering section the IR solitons are decelerated, and will therefore collide with the leading edge of the UV DWs, and thereby excite a non-degenerate FWM conversion deeper into the UV. Additionally, the increased dispersion and reduced core size in the tapering section facilitates a soliton-DW phase-matching condition further into the UV than with the original core size \cite{Ermolov-2015-PhysRevA.92.033821}. During multiple collisions in the tapering section we eventually have generated 2 XUV DWs with 190 nJ of total energy and DW1 centered as far down as 92 nm. As the light propagates in the waist section the propagation losses quickly dampens the UV and IR content, so it is important to propagate as little as possible after the up-conversion has finished. We should here as a practicality mention that we tried to put a reverse taper after the short waist section, and it seems to work quite well to reduce the propagation loss if a longer fiber is needed. 

More details into the dynamics can be found by studying the spectrograms, which are shown in Fig. \ref{fig:XFROG} at suitable stages of propagation: The first is right before the start of the taper, $z=0.17$ m, where two DWs have formed and the soliton fission process has commenced. After propagating 8 cm of the taper ($z=0.25$ m) the DW2 has been up-converted to the same wavelength as DW1 (140 nm) by soliton collision, and now 3 solitons are seen. After additional 5 cm of propagation ($z=0.30$ m) the solitons have collided with DW1 and DW2, so they are now up-converted to the XUV. At this stage 4 clear solitons are seen. Finally, at the taper stop ($z=0.36$ m) DW1 has been up-converted once again to 92 nm due to another soliton collision. The solitons are now quite weak, which is in turn due to ionization losses just before the taper stop as well as significant propagation losses at the waist core size. Note also that the final major ionization event has blue-shifted the 3. soliton, which is a process that may destroy the soliton. 

Already at the early soliton-fission stage ($z=0.17$ m) the resonances have left significant traces of dispersion delay in the spectrogram, as evidenced by the dotted gray line showing the accumulated dispersion delay $\tau(\omega,z)=[\beta_1(\omega,z)-\beta_1(\omega_0,z=0)]z$ where $\beta_1(\omega,z)$ is the frequency-dependent inverse group velocity and the $z$ dependence of it indicates that the dispersion changes in the tapering section. It is clear that the resonances are very dispersive, giving rise to quite sharp spectral lines and corresponding long pulses in the spectrogram. This notwithstanding, by comparing with a simulation using  the basic capillary MS model without the resonances of the poor-man's model, we find that in the nonlinear dynamics shown here is by and large unaffected by the dispersion of the resonances, while in turn the losses induced by the resonances and the glass material losses play a much larger role. This conclusion is quite sensitive to the resonance strength; in \cite{bache:2018} the original impedances used in the perturbative extension of the MS results were found to give inaccurate loss value in the peaks, i.e. at the resonances. The losses were a factor 4 larger than the BR calculations, and this, in turn, led to much stronger dispersion at the resonance. Essentially the dispersion delay in the resonances is much stronger and this significantly affects the broadband nonlinear interaction. 

\begin{figure}[t]
  \centering
  \includegraphics[width=0.49\linewidth]{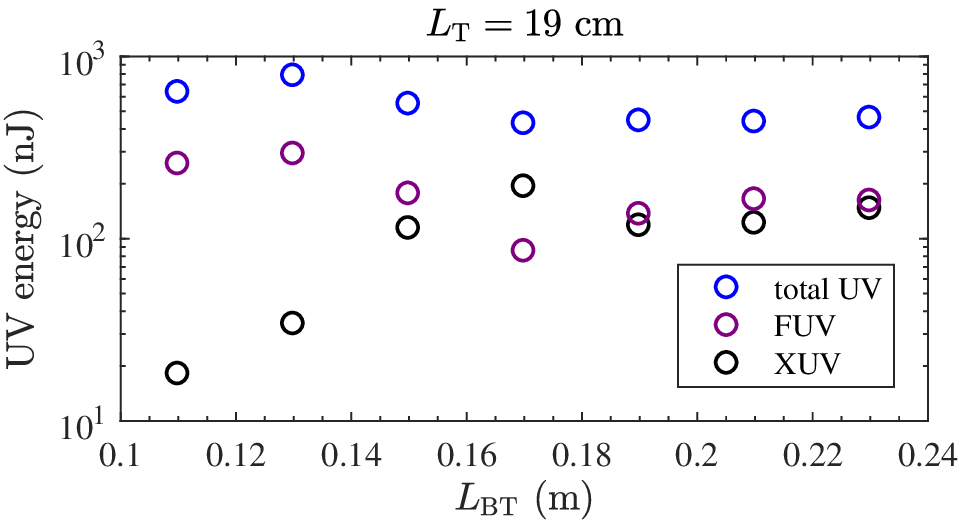}
  \includegraphics[width=0.49\linewidth]{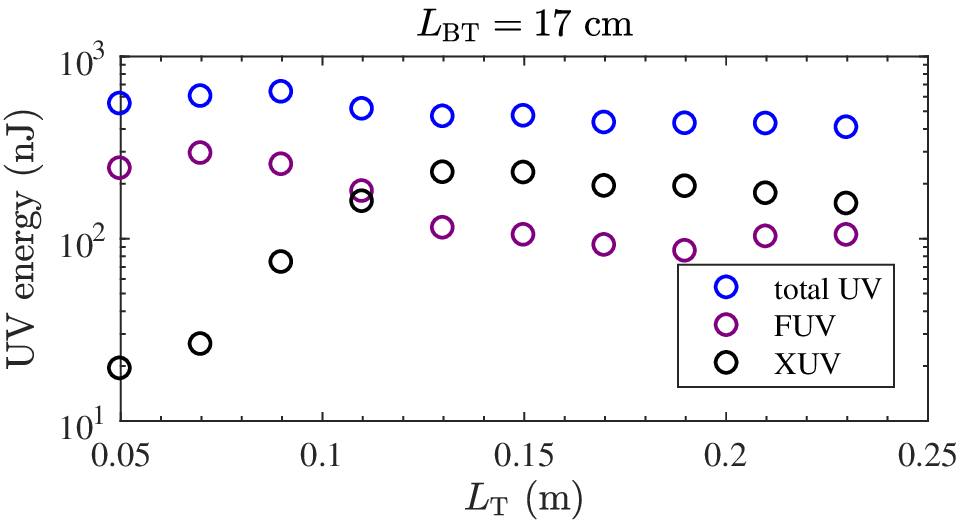}
  \includegraphics[width=0.49\linewidth]{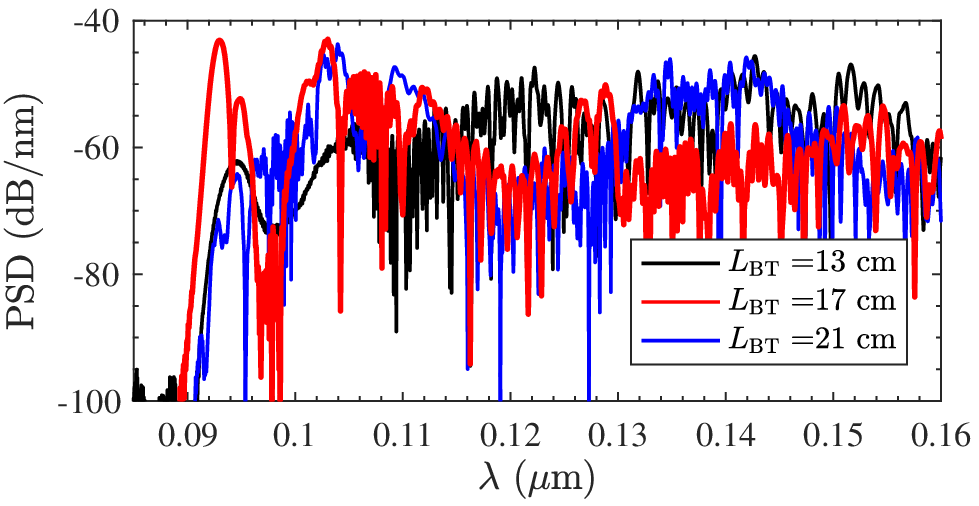}
  \includegraphics[width=0.49\linewidth]{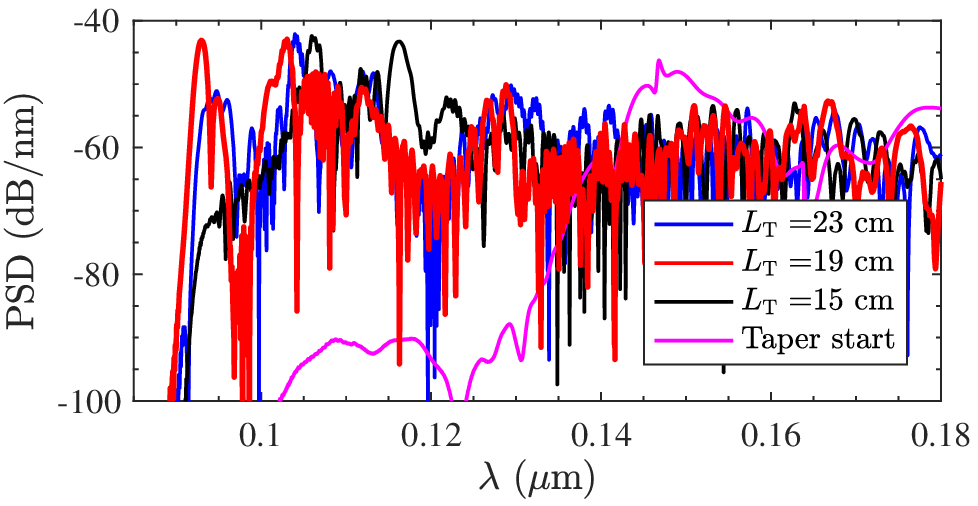}
\caption{UV energy and XUV part of selected spectra for two cases: fixed taper length (left) and fixed length before taper (right). The data are shown at the taper end, and do therefore not include propagation in the waist section of the taper. Note that in the two spectral plots, the red spectrum is the same, namely that of Fig. \ref{fig:PSD} and \ref{fig:XFROG}.}
%\vspace{-12pt}
\label{fig:taper}
\end{figure}

We investigated in detail the optimal taper shape. It can be made longer and shorter, and be introduced earlier or later in the soliton self-compression dynamics stage. A very short taper quickly decelerates the solitons and gives a short interaction length with the DWs for up-conversion. This limits the efficiency, but it can in certain cases be counter-balanced by a reduced propagation loss in the shorter taper. Starting the taper right at the self-compression stage turned out not to be ideal, most likely because the soliton fission stage is interrupted, and also because in many cases DW2 was not up-converted in the first stage to the same wavelength of DW1. Starting the taper too late leads to a too big delay between the solitons and the DWs and therefore a very long taper is needed to efficiently up-convert. A summary of these points is presented in Fig. \ref{fig:taper}, which shows variations around the optimized case we have presented so far in the previous simulations (17 cm before the taper, followed by 19 cm taper). 

If we start with a fixed taper length ($L_{\rm T}=19$ cm), we see that a very early taper start gives a high overall UV energy, but with limited XUV energy. At $L_{\rm BT}=13$ cm and below the FUV part has now been converted to XUV, as the collision of the DWs with the solitons becomes inefficient. Here 17 cm stands out as the maximum XUV energy, and looking at the spectra we see that this case has the sub-100 nm DW while the very early and very late taper start do not. 

For a fixed length before the taper ($L_{\rm BT}=17$ cm), we see the same trend: a very short taper gives only little XUV energy, while above 10 cm taper length the XUV yield is high, indicating that the DWs are upconverted. From the energy it does not appear that $L_{\rm T}=19$ cm is the optimal taper length, as the energy peaks at 13-15 cm. However, this does not take into account that only for the very long tapers do we get the sub-100 nm upconversion of DW1. This is seen in the spectra plotted below the energy figure. 

\begin{figure}[t]
  \centering
  \includegraphics[width=\linewidth]{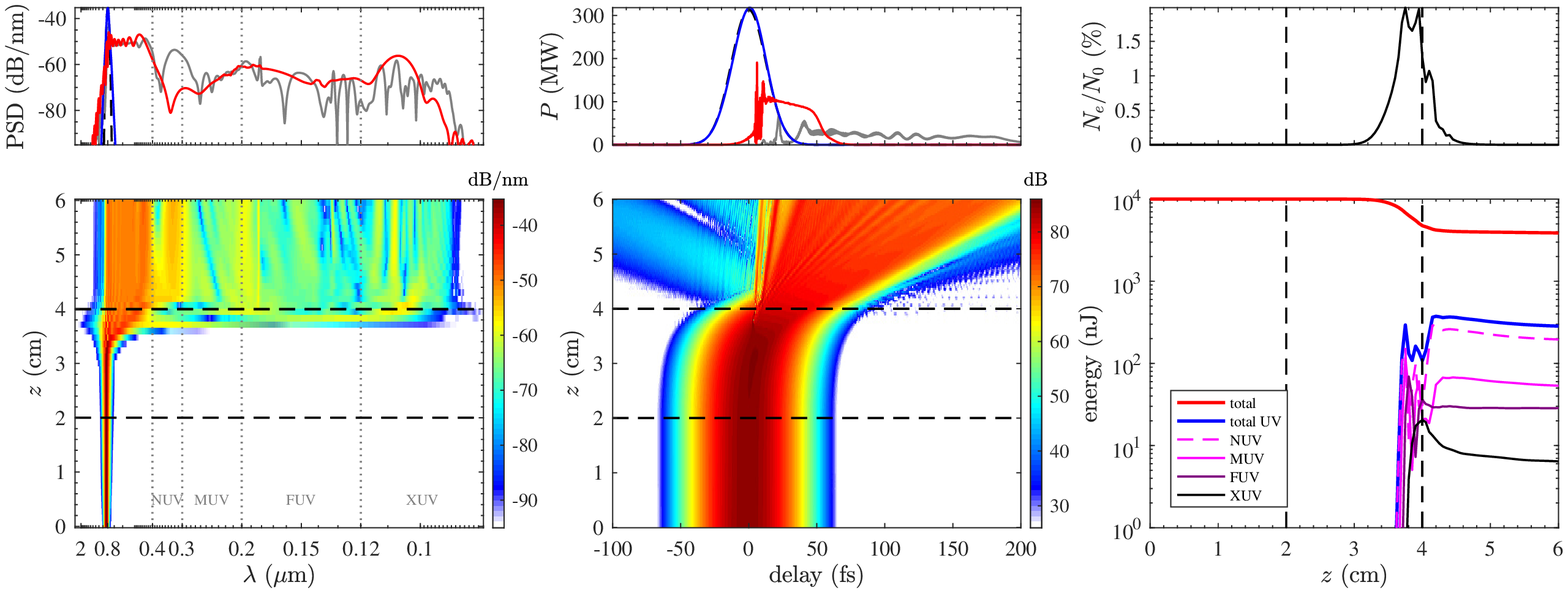}
  \includegraphics[width=0.49\linewidth]{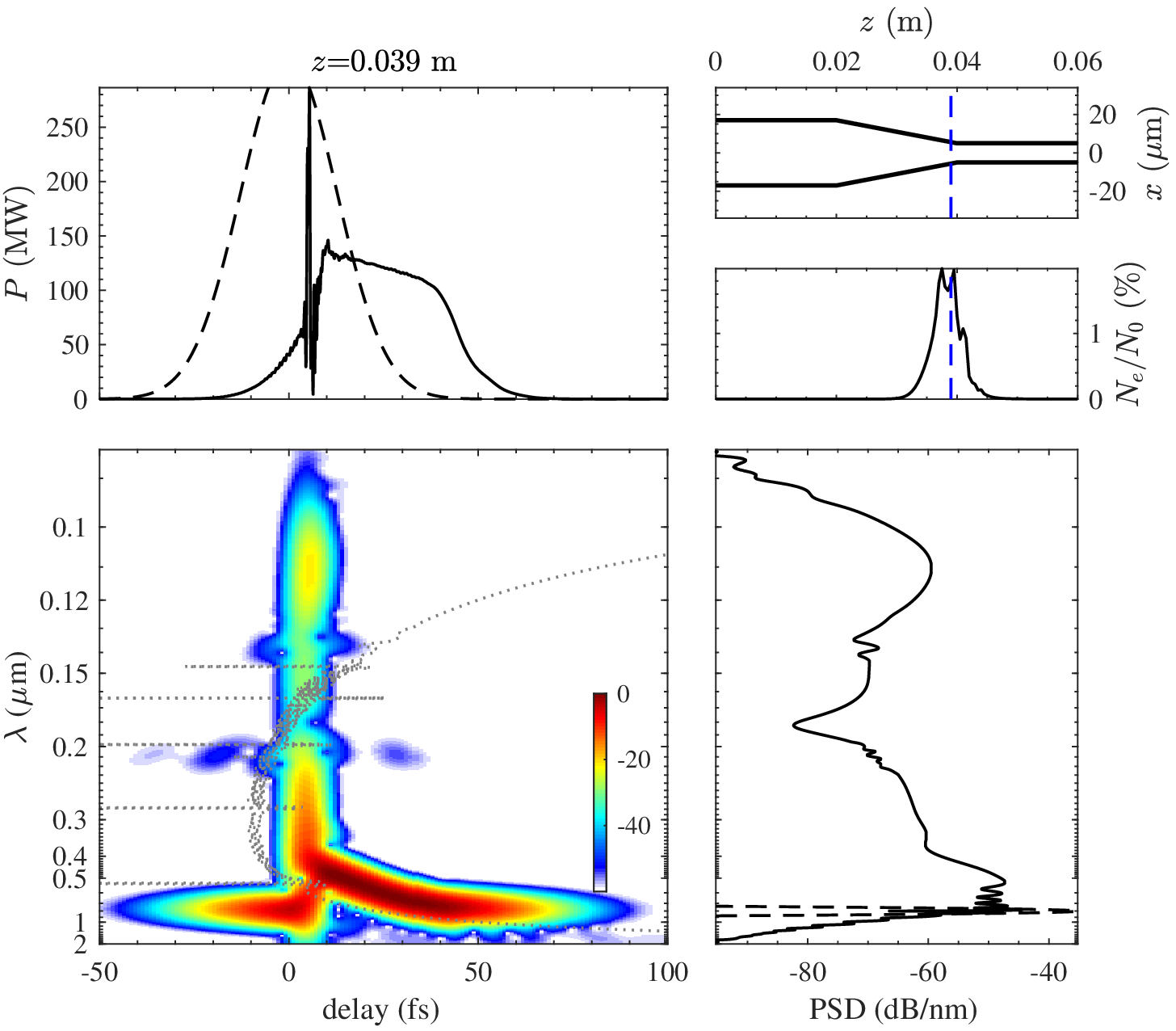}
  \includegraphics[width=0.49\linewidth]{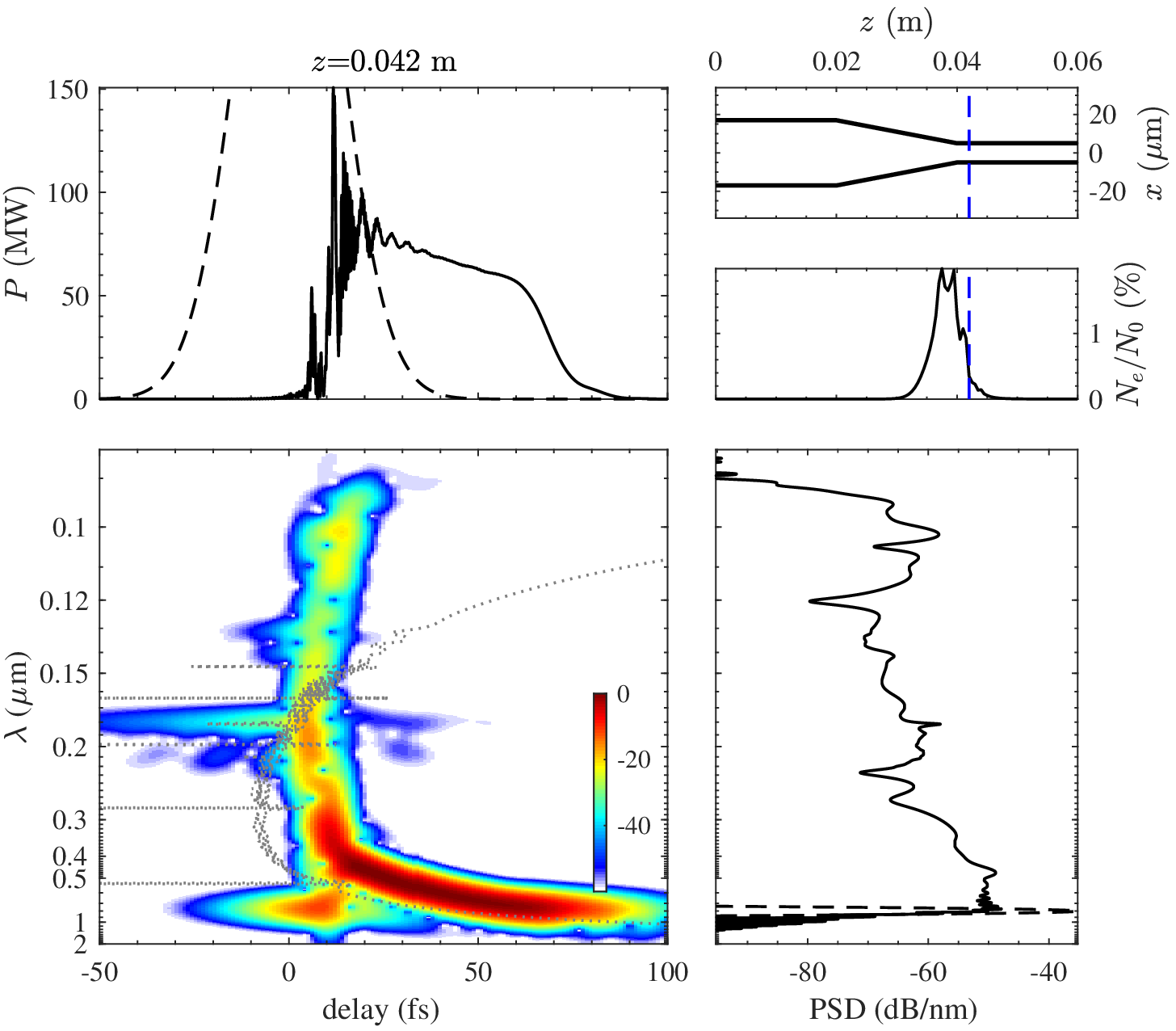}
\caption{Simulation of taper from 34 to $10\mic$ with extremely short section before the taper (2 cm) and taper section (2 cm). The dispersion and loss is as such the same as shown in Fig. \ref{fig:fiber_loss}, only the tapering section is much shorter. Again  a 800 nm 30 fs $10~\mu$J input pulse was used. The top row shows the spectral and temporal evolution vs. distance, as well as the ionization fraction and energy distribution. The bottom row shows selected spectrograms taken at the end of the taper.}
%\vspace{-12pt}
\label{fig:short}
\end{figure}

Finally, let us show an extreme example where the fiber taper is very short, both in the section before the taper starts and in the taper as such.  This case is shown in Fig. \ref{fig:short}, and since the taper sets in already after 2 cm, well in advance of the soliton self-compression stage, the result is that the spectrum broadens massively towards the blue during the taper and that at the taper exit a shock front forms at the leading edge. This generates a massive broadening towards the UV, with a peak centered around 105 nm in the XUV. This, unlike the previous soliton-case, is a 1-step conversion to the XUV. The energy is at the taper exit 15 nJ in the XUV. The spectrograms reveal that the dynamics seem quite identical to optical wave-breaking, but we remind that the tapering section does not introduce normal dispersion. On the contrary, strong anomalous dispersion is introduced to the pulse before the soliton can form, so what happens is that the dispersion length is reduced dramatically and GVD slows down the SPM-generated colors on the red side of the pump and pulls them away from the pulse center so the soliton cannot form. Eventually a very short spike forms on the leading edge, converting photons to the XUV. This is in part aided by the massive blue-shift of the pulse center due to ionization. 

\section{Conclusion}

In conclusion we have investigated a new approach to generate XUV coherent pulsed radiation in an HC-AR gas-filled fiber based on soliton-DW dynamics. By tapering a Ne-filled fiber we showed that the solitons could be slowed down to recollide with the UV DWs generated in the untapered stage. This up-converted the far-UV DWs to the extreme-UV, and we found that over 180 nJ of energy could be generated in the 90-120 nm range from $10~\mu$J 30 fs input at 800 nm. We also showed that a very short taper much before the soliton forms could perform a direct DW conversion to the extreme-UV, but with much less energy. 
Achieving DW generation below 110 nm has not been predicted before, and this we find can only be achieved with a tapered fiber. 
We here used much lower pressures (Ne, 9 bar) than other papers in the literature, where a high pressure is needed to achieve far- to extreme-UV phase matching. This lower pressure is more practical to implement experimentally and it also reduces ionization losses. We found similar dynamics also at lower pressures than 9 bar, but the challenge with lower pressures is to generate the first soliton compression stage in a 10-20 cm fiber without having to use too high pump energies. 

The simulations relied on a model recently developed for this purpose \cite{bache:2018}, where an analytical extension of the capillary model was used to accurately describe the anti-resonant and resonant transmission bands and how they affect the dispersion as well as loss. Importantly, the model also takes into account the lossy nature of glass, which is important in the UV. This analytical extension of the capillary model is a quick way of mimicking complicated and detailed finite-element simulations, so we expect it to find broad usage in the community. 

\end{document}